\documentclass[aps,prd]{revtex4}
\usepackage[colorlinks=true, pdfstartview=FitV, linkcolor=blue, citecolor=red, urlcolor=magenta]{hyperref}
\usepackage{graphicx}
\usepackage{latexsym}
\usepackage{amsmath}
\usepackage{amsfonts}
\usepackage{amssymb}
\usepackage{verbatim}

\newcommand{\be}{\begin{equation}}
\newcommand{\ee}{\end{equation}}
\newcommand{\bea}{\begin{eqnarray}}
\newcommand{\eea}{\end{eqnarray}}

\newcommand{\pa}{\partial}

\newcommand{\bb}{\bibitem}

\def\bb{\bibitem}
\def\as{A\!\!\!/}

\def\ns{n\!\!\!/}
\def\Asv{\mathcal{A}_{\mu}}
\def\TAsv{\tilde{\mathcal{A}}_{\mu}}

\def\As{\mathcal{A}\!\!\!/}
\def\TAs{\tilde{\mathcal{A}}\!\!\!/}
\def\Bs{\mathcal{A}\!\!\!\!/}

\def\Bsv{\mathcal{A}_{\mu}}
\def\TBsv{\tilde{\mathcal{A}}_{\mu}}
\def\Bsvv{\mathcal{A}^{\mu}}
\def\TBsvv{\tilde{\mathcal{A}}^{\mu}}
\def\BBs{\tilde{\mathcal{A}}\!\!\!/}

\def\ps{p\!\!\!/}
\def\bs{b\!\!\!/}

\def\ds{\partial\!\!\!/}

\def\bb{\bibitem}
\newcommand{\ben}{\begin{eqnarray}}
\newcommand{\een}{\end{eqnarray}}

\begin{document}

\title{Induction of the Lorentz-violating effective actions in quantum electrodynamics}

\author{$^{1}$M. A. Anacleto}
\email{anacleto@df.ufcg.edu.br}
\author{ $^{1,2}$F. A. Brito}
\email{fabrito@df.ufcg.edu.br}
\author{$^{3}$O.  Holanda}
\email{ozorio.neto@uerj.br}
\author{$^{1}$E. Passos}
\email{passos@df.ufcg.edu.br}

\affiliation{$^{1}$Departamento de F\'{\i}sica, Universidade Federal de Campina Grande,\\
Caixa Postal 10071, 58429-900, Campina Grande, Para\'{\i}ba, Brazil.}
\affiliation{$^{2}$Departamento de F\' isica, Universidade Federal da Para\' iba,\\  Caixa Postal 5008, Jo\~ ao Pessoa, Para\' iba, Brazil.}
\affiliation{$^{3}$Departamento de F\'{i}sica Te\' orica, Instituto de F\' isica, Universidade do Estado do Rio de Janeiro,
Rua S\~ ao Francisco Xavier 524,  20550-013,  Maracan\~ a,  Rio de Janeiro, Brazil}




\begin{abstract}
We introduce a Lorentz-symmetry violating extended quantum electrodynamics (QED) which preserves  gauge symmetry. 
The extended fermionic sector can radiatively induce an extended effective action which simultaneously displays the same electromagnetic terms present
in the Carroll-Field-Jackiw, Myers-Pospelov and Aether actions. 
\end{abstract}
\pacs{XX.XX, YY.YY} \maketitle


\section{Introduction}

Investigations consider the possibility of Lorentz-symmetry violation from a modification of the usual Standard Model of particle physics \cite{k01,k02,k03,k04}. This procedure is called the Standard Model Extension (SME) which provides a quantitative description of the Lorentz-symmetry violation, controlled by a set of coefficients whose values {\bf might} be determined or constrained by experiments \cite{explv,explv2,explv3,explv4}. Although, this approach introduces the Lorentz-symmetry violation it preserves all the conventional properties of quantum field theory such as gauge symmetry.

At the same time, the properties of the quantum corrections play a decisive role in establishing the main parameters associated to the structure of the SME. Here, we highlight interesting investigations such as radiative induction of Lorentz-symmetry violating electromagnetic effective actions.  This problem was first addressed in \cite{RC01,RC02,RC03,RC04,RC05,RC06,RC07,RC08}, in which was argued that the Carroll-Field-Jackiw (CFJ) model, ${\cal L}_{CFJ}\sim \varepsilon^{\mu\lambda\rho\nu}n_{\mu}F_{\lambda\rho}A_{\nu}$, with $n_{\mu}$ being a constant four-vector \cite{cfj, jackiw1}, can be radiatively induced from the Lorentz violation extension, $\bar\psi \bs\gamma_ {5}\psi$, in the fermionic sector, with a constant four-vector $b_{\mu}$ introducing CPT symmetry breaking. Among these developments one of the most interesting results is the relationship between the parameters $n_{\mu}\sim b_{\mu}$  present in the Lorentz and CPT symmetries breaking obtained when we integrate over the fermion fields in the modified Dirac action. Similar considerations have been addressed in the literature in several different contexts such as extended QED at finite temperature \cite{T. Finita,T. Finita2,T. Finita3,T. Finita4,T. Finita5}, non-Abelian QED \cite{nonabelian}, massless extended QED \cite{massless,massless2},  gravity \cite{grav,grav2,grav3} and nonminimal extended QED \cite{nonm,nonm2,nonm3}.  

{The main goal of this work is to consider the following nonminimal coupling between fermions and photons $- \tilde{q}\bar\psi \gamma^{\mu}\varepsilon_{\mu\nu\alpha\beta}n^{\nu}F^{\alpha\beta}\psi$ as an extension of Lorentz-violating QED and  study the problem of radiative induction to the photon sector. The associated Lagrangian with this modified Lorentz-violating QED is written as}
\bea\label{I01}
{\cal L}_{\psi}= \bar{\psi}\big(i\ds - m\big)\psi - \bar\psi \ns\gamma_{5}\psi - q \bar\psi\as\psi - \tilde{q}\bar\psi \gamma^{\mu}\varepsilon_{\mu\nu\alpha\beta}n^{\nu}F^{\alpha\beta}\psi,
\eea 
being $\tilde{q}$  a nonminimal coupling constant. The above Lagrangian represent one new possibility to introduce the (Lorentz invariance violation) LIV in an usual quantum field theory. Such extension was analyzed in \cite{cfjadp,cfjadp2,cfjadp3}, as a CPT-odd nonminimal coupling between fermions and the gauge field in the Dirac equation.  At the one-loop level, the last term in the Lagrangian, Eq.(\ref{I01}), can be induced as vertex correction to minimal SME such as an anomalous magnetic moment of electron \cite{ver1} --- see also \cite{ver2,ver22,ver23,ver24}.

Note that the above Lagrangian presents three distinct interactions: the first is mediated by a non-dynamical field $n_{\mu}$, the second is given by dynamical gauge field $A_{\mu}$ and the last term corresponds to an interaction between a 
non-dynamical field $n_{\mu}$ and dynamical electromagnetic strength field $F_{\mu\nu}$. Then at the quantum level, it can provide additional vertices to one-loop radiative corrections which can lead to some technical difficulties. Thus, in order to overcome this problem,  the Lagrangian, Eq.(\ref{I01}), is rewritten in a simpler form
\bea\label{cfj-psi3}
\tilde{\cal L}_{\psi}=\bar\psi\big(i\ds - m - q\tilde{q}\big(\As P_{R}- \TAs P_{L} \big)\big)\psi,
\eea
where $P_{R}$ and $P_{L}$ are right- and left-handed projection operators $P_R = (1 + \gamma_{5})/2$ and $P_L = (1-\gamma_{5})/2$ and
\begin{subequations}
\bea\label{Ef1-21}
\Asv=\frac{1}{q\tilde{q}} n_{\mu}+ \frac{1}{\tilde{q}}A_{\mu} + \frac{1}{q}V_{\mu} ,
\eea
\bea\label{Ef2-22}
\TAsv=\frac{1}{q\tilde{q}} n_{\mu} - \frac{1}{\tilde{q}}A_{\mu} - \frac{1}{q}V_{\mu}  
\eea
\end{subequations}
where $V_{\mu} = \varepsilon_{\mu\nu\alpha\beta}n^{\nu}F^{\alpha\beta}$ is a LIV dynamical field. In this configuration we have that the quantities $\Asv$ and $\TAsv$ can be interpreted as  {\it bumblebee-like} fields (a LIV non-dynamical field perturbed by photon fields \cite{bumblebee}). 
Note also that the Lagrangian, Eq.(\ref{cfj-psi3}) written in terms of the projection operators $P_{R}$ and $P_{L}$ with the properties $P_{R}\times P_{L}=P_{L}\times P_{R}=0$, can play a differential role in the one-loop quantum calculation. 

In this paper, we shall take this into consideration by using the expressions given by Eq.(\ref{cfj-psi3}) and Eqs.(\ref{Ef1-21}) -- (\ref{Ef2-22}) to address the specific problem of radiatively inducing some LIV effective actions such as CPT-odd Chern-Simons-like \cite{cfj, jackiw1}, CPT-even aether-like \cite{petrov2} and CPT-odd extensions with higher order derivatives \cite{MP01, MP, sln, grbnossos}, given respectively as 
\bea\label{livs}
&&{S}_{CS} \sim \int d^{4}x\, \varepsilon^{\alpha\lambda\mu\nu}n_{\alpha}F_{\lambda\mu}A_{\nu},
\;\;\;
{S}_{AET} \sim \int d^{4}x\, n^{\beta} n_{\alpha}F^{\lambda\alpha} F_{\lambda\beta},
\;\;{\rm and}\nonumber\\&&
{ S}_{ECS} \sim \int d^{4}x\, n^{\alpha} F_{\alpha\mu} \big(n\cdot\pa\big) \big(n_{\beta}\tilde{F}^{\beta\mu}\big)
\eea
respecting the gauge conditions: $\pa\cdot A = n\cdot A=0$.
The effective actions given by Eq.(\ref{livs}) can be obtained from the general effective actions of the following types:
\begin{subequations}
\bea\label{effa1}
{S}^{(2)}_{gen}\sim q^{2}\tilde{q}^{2} \int d^{4}x \Big( c_{1}\, \Bsv \Bsvv + c_{2}\, \Bsv\TBsvv+ c_{3}\,  \TBsv\TBsvv\Big) 
\eea
\bea\label{effa2}
{S}^{(3)}_{gen}\sim q^{3}\tilde{q}^{3} \int d^{4}x \varepsilon^{\alpha\beta\mu\nu}\Big( c_{4}(\pa_{\beta}\tilde{\mathcal{A}}_{\mu})\tilde{\mathcal{A}}_{\alpha}\mathcal{A}_{\nu} + c_{5}(\pa_{\beta}\mathcal{A}_{\mu}){\mathcal{A}}_{\alpha}\tilde{\mathcal{A}}_{\nu} \Big)
\eea
\end{subequations}
where $c_{n}$ with $n=1,2,3,...$ are numerical constants. We will address the specific problem of radiatively inducing the Eq.(\ref{effa1}) and Eq.(\ref{effa2}) as radiative corrections at one loop. Therefore, we shall show that is possible to generate the previous LIV effective actions, Eq.(\ref{livs}), which may offer important contributions to the usual electrodynamics.

The structure of the work is organized as follows: In the section \ref{sec01} we construct a general effective action from fermion determinant associated with the extended QED given by Eq.(\ref{cfj-psi3}) and we isolate the self-energy tensor via derivative expansion method. In the section \ref{sec02} we calculate the momentum integrals from dimensional regularization scheme.
In the section \ref{sec03}, the results show that the effective action can be generated and offer high derivative Lorentz-violating operators which may be added into the usual electrodynamics for future investigations. In section \ref{sec04} we present our conclusions. Some useful formulas associate to dimensional regularization scheme are presented in Appendix \ref{APPA}.



\section{The Effective Action: fermionic determinant}\label{sec01}
Let us now start the process of radiative induction of the effective actions, Eq.(\ref{effa1}) and Eq.(\ref{effa2}) by using the derivative expansion of fermion determinants.  Thus, we consider a non-minimal Lagrangian, Eq.(\ref{cfj-psi3}) and to taking into account  the fermionic integration we write
\bea\label{aa1}
e^{iS_{eff}}=\int D\bar{\psi}D\psi\, e^{i\int d^{4}x {\tilde{\cal L}_{\psi}}}
\eea
where the effective action is given by
\bea\label{aa4}
S_{eff}[\mathcal{A}_{\mu}, \tilde{\mathcal{A}}_{\mu}]= -i{\rm\, Tr\, ln}\big[\ps - m- q \tilde{q}\big(\Bs P_{R}- \BBs P_{L} \big)\big]
\eea
where the symbol $\rm Tr$ stands for the trace over Dirac matrices, trace over the internal space
as well as for the integrations in momentum and coordinate spaces.
The first nondynamical determinant factor has been absorbed into normalization. {Using the logarithm property one can split the Eq.~(\ref{aa4}) into two parts:
\bea\label{031}
S_{eff}[\mathcal{A}_{\mu}, \tilde{\mathcal{A}}_{\mu}] = S_{eff} + S_{eff}[\mathcal{A}_{\mu}, \tilde{\mathcal{A}}_{\mu}].
\eea
where the first term is  $S_{eff}= - i {\rm\, Tr\, ln} \big[ \ps - m \big]$, which does not depend on the effective fields. 
And the second term of Eq.~(\ref{031}) depends on the effective fields and is written in the form}
\bea\label{04}
S_{eff}[\mathcal{A}_{\mu}, \tilde{\mathcal{A}}_{\mu}] &=& - i {\rm Tr\, ln} \big[ 1 - S(p)  \big(\Bs\, P_{R}- \BBs\,P_{L} \big)\big]\nonumber\\&=&
-i{\rm Tr}\sum^{\infty}_{l=1}\frac{1}{l}\Big[-i q\tilde{q}\, S(p) \big(\Bs\, P_{R}- \BBs\,P_{L} \big)\Big]^{l},
\eea
where 
\bea\label{prop}
S(p)=\frac{i}{\ps - m}
\eea 
is the  free fermion propagator. {All the terms of the series, Eq.~(\ref{04}), are all of one-loop contributions that generate the effective actions of the type given in Eqs.(\ref{effa1})--(\ref{effa2}). Before making explicit calculations, one can show that the possible non-zero terms are given in the general form as follows. 
First to $l=j_{1} - 2$ (for even and integer numbers $j_{1} > 2$) it is possible to obtain the Eq.(\ref{effa1}) in the form
\bea\label{effa1aa}
{S}^{(2)}_{gen}\sim q^{2}\tilde{q}^{2} \int d^{4}x \Big(c_{1} \Bsv \Bsvv + c_{2} \Bsv\TBsvv+ c_{3} \TBsv\TBsvv\Big)\bigg( 1+  \sum_{j_{1}} \frac{\bar{c}_{j_{1}(j_{1}-2)}}{(m^{2})^{(j_{1}-2)}}\Big({n^{2}}\Big)^{j_{1}-2}  \bigg).
\eea
And to $l=j_{2} - 3$ (for odd and integer numbers $j_{3} > 3$) one obtains Eq.~(\ref{effa2}) as	
\bea\label{effa2a}
{S}^{(3)}_{gen}\sim q^{3}\tilde{q}^{3} \int d^{4}x \varepsilon^{\alpha\beta\mu\nu}\Big( (c_{4}\pa_{\beta}\tilde{\mathcal{A}}_{\mu})\tilde{\mathcal{A}}_{\alpha}\mathcal{A}_{\nu} + c_{5}(\pa_{\beta}\mathcal{A}_{\mu}){\mathcal{A}}_{\alpha}\tilde{\mathcal{A}}_{\nu} \Big)\Big(1 + \sum_{j_{2}}\bar{c}_{j_{2}} (j_{2}-3)\bigg(\sqrt{\frac{n^{2}}{m^{2}}}\Big)^{j_{2}-3}\bigg).
\eea
where $\bar{c}_{j_{1}}$ and $\bar{c}_{j_{2}}$ are numerical constants.
Therefore, we observe that the terms associated with the Eq.(\ref{04}) for $l=4,5,...$ only correct the coupling constant and thus does not affect the dynamics of the induced effective actions. Thus, it suffices to restrict the induction of such effective actions to lower orders.
}

In the following we shall focus only on the term  associate to $l=2$ and $l=3$. Thus, we rewrite this contribution as
\bea\label{effm}
S_{eff}[\mathcal{A}_{\mu}, \tilde{\mathcal{A}}_{\mu}]=S_{eff}^{(2)} + S_{eff}^{(3)}+ \cdot\cdot\cdot
\eea
with,
\begin{subequations}
\bea\label{07a}
S_{eff}^{(2)}=\frac{i}{2} q^{2}\tilde{q}^{2}{\rm Tr}\big[ S(p)\Bs P_{R} S(p)\Bs P_{R} - 2 S(p)\Bs P_{R} S(p)\BBs P_{L} + S(p)\BBs P_{L} S(p)\BBs P_{L}  \big]
\eea
\bea\label{07b}
S_{eff}^{(3)} &=&-\frac{1}{3} q^{3} \tilde{q}^{3}{\rm Tr}\big[3 S(p)\Bs P_{R} S(p)\Bs P_{R} S(p)\BBs P_{L} - 3 S(p)\BBs P_{L} S(p)\BBs P_{L}  S(p)\Bs P_{R} - \nonumber\\&& S(p)\Bs P_{R} S(p)\Bs P_{R} S(p)\Bs P_{R}  +
S(p)\BBs P_{L} S(p)\BBs P_{L}S(p)\BBs P_{L} \big].
\eea
\end{subequations}
{
The Eq.(\ref{07a}) will induce the effective action Eq.(\ref{effa1}), whereas the Eq.(\ref{07b}) will induce the effective action Eq.(\ref{effa2}). 
}
 
Now we apply the main property of derivative expansion method \cite{expd,expd2,expd3,expd4,expd5,expd6}. This method consider that any function of momentum can be converted into a coordinate dependent quantity as
\bea\label{08}
(\mathcal{A}_{\mu}, \tilde{\mathcal{A}}_{\mu})S(p) = \big(S(p - i \pa)(\mathcal{A}_{\mu}, \tilde{\mathcal{A}}_{\mu})\big),
\eea
where
\bea\label{09}
\big(S(p - i \pa)(\mathcal{A}_{\mu}, \tilde{\mathcal{A}}_{\mu})\big)= S(p) (\mathcal{A}_{\mu}, \tilde{\mathcal{A}}_{\mu}) + S(p)\gamma^{\lambda} S(p) \big(\pa_{\lambda} (\mathcal{A}_{\mu}, \tilde{\mathcal{A}}_{\mu})\big) +\cdot\cdot\cdot
\eea
{ In the next step we shall insert the properties (\ref{08}) -- (\ref{09}) into Eqs.(\ref{07a}) -- (\ref{07b}) to appropriately induce the effective actions Eqs.(\ref{effa1})--(\ref{effa2}). To this end, we shall keep only the zero order terms in the derivative expansion applied in ${S^{(2)}_{eff}}$ to reproduce the Eq.(\ref{effa1}) and only the first order terms in the derivative expansion applied in ${S^{(3)}_{eff}}$ to reproduce the Eq.(\ref{effa2}). This leads us to
}
\bea\label{effb1}
S_{eff}^{(2)}=q^{2}\tilde{q}^{2} \int d^{4}x \Big( \mathcal{A}_{\mu} \mathcal{A}_{\nu} I^{\mu\nu}_{1} - 2\mathcal{A}_{\mu} \tilde{\mathcal{A}}_{\nu} I^{\mu\nu}_{2} + \tilde{\mathcal{A}}_{\mu}\tilde{\mathcal{A}}_{\nu} I^{\mu\nu}_{3}\Big) 
\eea
being $I^{\mu\nu}_{1,2,3}$ given as
\begin{subequations}
\bea\label{tp01}
I^{\mu\nu}_{1}= \frac{i}{2} \int \frac{d^{4}p}{(2\pi)^{4}} {\rm tr}\big[ S(p)\gamma^{\mu} P_{R} S(p)\gamma^{\nu} P_{R}\big]
\eea
\bea\label{tp02}
I^{\mu\nu}_{2}= \frac{i}{2} \int \frac{d^{4}p}{(2\pi)^{4}} {\rm tr}\big[ S(p)\gamma^{\mu} P_{R} S(p)\gamma^{\nu} P_{L}\big],
\eea
\bea\label{tp03}
I^{\mu\nu}_{3}=\frac{i}{2}  \int \frac{d^{4}p}{(2\pi)^{4}} {\rm tr}\big[ S(p)\gamma^{\mu} P_{L} S(p)\gamma^{\nu} P_{L}\big].
\eea
\end{subequations}
And then,
\bea\label{10}
S_{eff}^{(3)}&=& - q^{3} \tilde{q}^{3}\int d^{4}x \Big[ 3 \Big((\pa_{\lambda}\mathcal{A}_{\mu}) \mathcal{A}_{\nu}\tilde{\mathcal{A}}_{\alpha}\,I^{\mu\lambda\nu\alpha}_{1} -   (\pa_{\lambda}\tilde{\mathcal{A}}_{\mu})\tilde{\mathcal{A}}_{\nu}\mathcal{A}_{\alpha}\,I^{\mu\lambda\nu\alpha}_{2}\Big) - \nonumber\\&&
\Big((\pa_{\lambda}\mathcal{A}_{\mu}) \mathcal{A}_{\nu}{\mathcal{A}}_{\alpha}\,I^{\mu\lambda\nu\alpha}_{3} -
(\pa_{\lambda}\tilde{\mathcal{A}}_{\mu})\tilde{\mathcal{A}_{\nu}}\tilde{\mathcal{A}}_{\alpha}\,I^{\mu\lambda\nu\alpha}_{4}\Big)\Big]
\eea
with $I^{\mu\lambda\nu\alpha}_{1,2,3,4}$ written as 
\begin{subequations}
\bea\label{11}
I^{\mu\lambda\nu\alpha}_{1}&=& \frac{1}{3}\int \frac{d^{4}p}{(2\pi)^{4}}{\rm tr}\big[S(p)\gamma^{\mu} P_{R}S(p)\gamma^{\lambda} S(p)\gamma^{\nu}P_{R}S(p)\gamma^{\alpha}P_{L}+S(p)\gamma^{\nu} P_{R}S(p)\gamma^{\mu}P_{R} S(p)\gamma^{\lambda}S(p)\gamma^{\alpha}P_{L}+\nonumber\\&& \;\;\;\;\;\;\;\;\;\;\;\;\;\;\;\;\;S(p)\gamma^{\mu} P_{R}S(p)\gamma^{\nu}P_{R} S(p)\gamma^{\lambda}S(p)\gamma^{\alpha}P_{L}\big],
\eea
\bea\label{12}
I^{\mu\lambda\nu\alpha}_{2}&=& \frac{1}{3}\int \frac{d^{4}p}{(2\pi)^{4}}{\rm tr}\big[S(p)\gamma^{\mu} P_{L}S(p)\gamma^{\lambda} S(p)\gamma^{\nu}P_{L}S(p)\gamma^{\alpha}P_{R}+S(p)\gamma^{\nu} P_{L}S(p)\gamma^{\mu}P_{L} S(p)\gamma^{\lambda}S(p)\gamma^{\alpha}P_{R}+\nonumber\\&& \;\;\;\;\;\;\;\;\;\;\;\;\;\;\;\;\;S(p)\gamma^{\mu} P_{L}S(p)\gamma^{\nu}P_{L} S(p)\gamma^{\lambda}S(p)\gamma^{\alpha}P_{R}\big],
\eea
\bea\label{11a}
I^{\mu\lambda\nu\alpha}_{3}&=& \frac{1}{3}\int \frac{d^{4}p}{(2\pi)^{4}}{\rm tr}\big[S(p)\gamma^{\mu} P_{R}S(p)\gamma^{\lambda} S(p)\gamma^{\nu}P_{R}S(p)\gamma^{\alpha}P_{R}+S(p)\gamma^{\nu} P_{R}S(p)\gamma^{\mu}P_{R} S(p)\gamma^{\lambda}S(p)\gamma^{\alpha}P_{R}+\nonumber\\&& \;\;\;\;\;\;\;\;\;\;\;\;\;\;\;\;\;S(p)\gamma^{\mu} P_{R}S(p)\gamma^{\nu}P_{R} S(p)\gamma^{\lambda}S(p)\gamma^{\alpha}P_{R}\big],
\eea
\bea\label{12b}
I^{\mu\lambda\nu\alpha}_{4}&=& \frac{1}{3}\int \frac{d^{4}p}{(2\pi)^{4}}{\rm tr}\big[S(p)\gamma^{\mu} P_{L}S(p)\gamma^{\lambda} S(p)\gamma^{\nu}P_{L}S(p)\gamma^{\alpha}P_{L}+S(p)\gamma^{\nu} P_{L}S(p)\gamma^{\mu}P_{L} S(p)\gamma^{\lambda}S(p)\gamma^{\alpha}P_{L}+\nonumber\\&& \;\;\;\;\;\;\;\;\;\;\;\;\;\;\;\;\;S(p)\gamma^{\mu} P_{L}S(p)\gamma^{\nu}P_{L} S(p)\gamma^{\lambda}S(p)\gamma^{\alpha}P_{L}\big].
\eea
\end{subequations}
{ In this way, we have that the Eqs.(\ref{tp01})--(\ref{tp03}) represent the possible self-energy tensors at zero order derivative and the Eqs.(\ref{11})--(\ref{12b}) represent the possible self-energy tensors at first order derivative. Notice also that the others contributions at first order derivative into Eq.(\ref{07a}) and zero order derivative into Eq.(\ref{07b}) give rise to terms that are of odd exponent at $p$-momentum which must be null from an integration over the symmetric interval.}
Here, the symbol $\rm tr$ denotes the trace of the product of the gamma matrices. 

\section{Calculating the Momentum Integrals}\label{sec02}
At this point, we will calculate the integrals associated with the above self-energy tensors, Eqs.(\ref{tp01}) -- (\ref{12b}). From calculations of trace on gamma matrices, we can group the results obtained into three sets of momentum integrals. The first set of integrals is given as
\begin{subequations}
\bea\label{tns01}
I^{\mu\nu}_{1}=I^{\mu\nu}_{3}=  i\int\, \frac{d^{4}p}{(2\pi)^{4}} \frac{p^{2} \eta^{\mu\nu} - 2p^{\mu}p^{\nu} }{(p^{2} - m^{2})^{2}},
\eea
\bea\label{tns02}
I^{\mu\nu}_{2}=-i \int\, \frac{d^{4}p}{(2\pi)^{4}} \frac{m^{2} \eta^{\mu\nu}}{(p^{2} - m^{2})^{2}}.
\eea
\end{subequations}
The second set of integrals can be written as $I^{\mu\lambda\nu\alpha}_{1} \to I^{\mu\lambda\nu\alpha}_{1.a}+I^{\mu\lambda\nu\alpha}_{1.b}$ and $I^{\mu\lambda\nu\alpha}_{2} \to I^{\mu\lambda\nu\alpha}_{2.a}+I^{\mu\lambda\nu\alpha}_{2.b}$ where
\begin{subequations}
\bea\label{tns03}
I^{\mu\lambda\nu\alpha}_{1.a}=I^{\mu\lambda\nu\alpha}_{2.a}&=& \frac{2}{3}m^{2} \int \frac{d^{4}p}{(2\pi)^{4}}\,\frac{(p^{2} - m^{2})(\eta^{\alpha\lambda}\eta^{\mu\nu}-\eta^{\alpha\mu}\eta^{\lambda\nu}-\eta^{\lambda\mu}\eta^{\alpha\nu})}{(p^{2} - m^{2})^{4}}+\nonumber\\
&&\frac{6(p^{\lambda}p^{\mu}\eta^{\alpha\nu}-p^{\alpha}p^{\lambda}\eta^{\mu\nu}+p^{\lambda}p^{\nu}\eta^{\alpha\mu})}{(p^{2} - m^{2})^{4}},
\eea
and
\bea\label{tns04}
I^{\mu\lambda\nu\alpha}_{1.b}=-I^{\mu\lambda\nu\alpha}_{2.b}= \frac{2i}{3}  m^{2} \int \frac{d^{4}p}{(2\pi)^{4}}\,\frac{2\varepsilon^{\beta\mu\alpha\nu}p_{\beta}p^{\lambda} - 
(p^{2} - m^{2})\varepsilon^{\lambda\mu\alpha\nu} }{(p^{2} - m^{2})^{4}}.
\eea
Finally, the following integrals can be written as $I^{\mu\lambda\nu\alpha}_{3} \to I^{\mu\lambda\nu\alpha}_{3.a}+I^{\mu\lambda\nu\alpha}_{3.b}$ and $I^{\mu\lambda\nu\alpha}_{4} \to I^{\mu\lambda\nu\alpha}_{4.a}+I^{\mu\lambda\nu\alpha}_{4.b}$ where
\bea\label{tns05}
I^{\mu\lambda\nu\alpha}_{3.a}=I^{\mu\lambda\nu\alpha}_{4.a}&=& \frac{2}{3}  \int \frac{d^{4}p}{(2\pi)^{4}}
\frac{2m^{2}(2p^{\mu}p^{\alpha}\eta^{\lambda\nu}-p^{\lambda}p^{\nu}\eta^{\mu\alpha}+2p^{\nu}p^{\mu}\eta^{\alpha\lambda}-p^{\alpha}p^{\lambda}\eta^{\nu\mu}+2p^{\nu}p^{\alpha}\eta^{\lambda\mu}-p^{\lambda}p^{\mu}\eta^{\nu\alpha})}{(p^{2} - m^{2})^{4}} - \nonumber\\&& \frac{4p^{2}(p^{\lambda}p^{\mu}\eta^{\alpha\nu}+p^{\alpha}p^{\lambda}\eta^{\mu\nu}+p^{\alpha}p^{\mu}\eta^{\lambda\nu}+p^{\nu}p^{\alpha}\eta^{\lambda\mu}+p^{\nu}p^{\lambda}\eta^{\alpha\mu}+p^{\nu}p^{\mu}\eta^{\alpha\lambda})}{(p^{2} - m^{2})^{4}}+\nonumber\\
&&\frac{p^{2}(p^{2}-m^{2})(\eta^{\alpha\nu}\eta^{\lambda\mu}+\eta^{\alpha\mu}\eta^{\lambda\nu}+\eta^{\alpha\lambda}\eta^{\mu\nu})+24p^{\mu}p^{\lambda}p^{\nu}p^{\alpha}}{(p^{2} - m^{2})^{4}},
\eea
and
\bea\label{tns06}
I^{\mu\lambda\nu\alpha}_{3.b}&=&-I^{\mu\lambda\nu\alpha}_{4.b}= \frac{2i}{3}  \int \frac{d^{4}p}{(2\pi)^{4}}
\frac{p^{2}(p^{2} - m^{2})  \varepsilon^{\alpha\lambda\mu\nu}-p^{2}p^{\lambda}p_{\beta}\varepsilon^{\alpha\mu\nu\beta}  }{(p^{2} - m^{2})^{4}} + \nonumber\\&& \frac{ 2 (p^{2} - m^{2}) \big( p^{\mu}p_{\beta}\varepsilon^{\alpha\lambda\nu\beta} + p^{\nu}p_{\beta}\varepsilon^{\alpha\lambda\mu\beta} + p^{\alpha}p_{\beta}\varepsilon^{\lambda\mu\nu\beta}\big)}{(p^{2} - m^{2})^{4}}.
\eea
\end{subequations}
Now, performing an analysis on the Eqs.(\ref{tns01}) - (\ref{tns06}) by momentum power counting we can see that the calculations involve finite integrals and others that diverge logarithmically and quadratic. We shall now use a regularization scheme that preserves gauge invariance such as the dimensional regularization scheme \cite{RegD} (see the main results in the Appendix \ref{APPA}). According to this scheme, the above integrals reduce to
\begin{subequations}
\bea\label{tnss01}
I^{\mu\nu}_{1}=I^{\mu\nu}_{3}=I^{\mu\nu}_{2} = 
\frac{m^{2}}{(4\pi)^2} \alpha(\epsilon) \Gamma\Big( \frac{\epsilon}{2}\Big)  \eta^{\mu\nu},
\eea
\bea\label{tnss02}
I^{\mu\lambda\nu\alpha}_{1.a}= I^{\mu\lambda\nu\alpha}_{2.a} =0\,,
\eea
\bea\label{tnss03}
I^{\mu\lambda\nu\alpha}_{1.b}= -I^{\mu\lambda\nu\alpha}_{2.b} =
-\frac{1}{(4\pi)^2}\frac{2}{9}\alpha(\epsilon)\,\varepsilon^{\mu\lambda\nu\alpha}\,,
\eea
\bea\label{tnss04}
I^{\mu\lambda\nu\alpha}_{3.a}= I^{\mu\lambda\nu\alpha}_{4.a} =0\,,
\eea
\bea\label{tnss05}
I^{\mu\lambda\nu\alpha}_{3.b}= -I^{\mu\lambda\nu\alpha}_{4.b} =-
\frac{1}{(4\pi)^2} \frac{1}{3}\Big(\frac{1}{2} - \frac{5\epsilon}{12} \Big) \alpha(\epsilon) \Gamma\Big(\frac{\epsilon}{2}\Big)\varepsilon^{\mu\lambda\nu\alpha}.
\eea
\end{subequations}

\section{The Induced Effective Actions}\label{sec03}
Let us now write the induced effective actions in an explicit form in terms of the gauge field. Thus, by inserting Eq.(\ref{tnss01}) into Eq.(\ref{effb1}) we obtain
\bea\label{27}
S_{eff}^{(2)}&=& \frac{C_{1}}{4} (q^{2}\tilde{q}^{2})\int d^{4}x \Big( \mathcal{A}_{\mu} \mathcal{A}^{\mu}  - 2\mathcal{A}_{\mu} \tilde{\mathcal{A}}^{\mu}  + \tilde{\mathcal{A}}_{\mu}\tilde{\mathcal{A}}^{\mu} \Big)
\eea
where 
\bea\label{27.a}
C_{1} =  \frac{ m^{2}}{4\pi^2} \alpha(\epsilon) \Gamma\Big( \frac{\epsilon}{2}\Bigl) . 
\eea
Note that $C_1$ into Eq.(\ref{27.a}) diverges in the limit $D\to4$. 
However, for massless photons $(m\rightarrow 0)$, this constant is zero and consequently the effective action given in Eq.(\ref{27}) also disappears. On the other hand, by inserting equations (\ref{tnss02}) -- (\ref{tnss05}) into Eq.(\ref{10}) we find (with $\varepsilon^{\mu\lambda\nu\alpha}
\big((\pa_{\lambda}\mathcal{A}_{\mu}) \mathcal{A}_{\nu}{\mathcal{A}}_{\alpha} +
(\pa_{\lambda}\tilde{\mathcal{A}}_{\mu})\tilde{\mathcal{A}_{\nu}}\tilde{\mathcal{A}}_{\alpha}\big) = 0$):
\bea\label{28}
S_{eff}^{(3)}&=&- \frac{( 3\, C_{2})}{4}(q^{3} \tilde{q}^{3}) \int d^{4}x \,\varepsilon^{\mu\lambda\nu\alpha} \big((\pa_{\lambda}\mathcal{A}_{\mu}) \mathcal{A}_{\nu}\tilde{\mathcal{A}}_{\alpha} + (\pa_{\lambda}\tilde{\mathcal{A}}_{\mu})\tilde{\mathcal{A}}_{\nu}\mathcal{A}_{\alpha} \big) 
\eea
where
\bea\label{28.a}
C_{2} = \frac{ 1}{2\pi^2}\frac{1}{9} \alpha(\epsilon).
\eea
In the limit $D\to 4$ we have only finite contribution.

Now, we are ready to examine in detail the induced effective action given in Eq.(\ref{28}) by considering the fields defined in Eq.(\ref{Ef1-21}) and (\ref{Ef2-22}). In doing this, we get
\bea\label{29.1}
\varepsilon^{\mu\lambda\nu\alpha} \big((\pa_{\lambda}\mathcal{A}_{\mu}) \mathcal{A}_{\nu}\tilde{\mathcal{A}}_{\alpha} + (\pa_{\lambda}\tilde{\mathcal{A}}_{\mu})\tilde{\mathcal{A}}_{\nu}\mathcal{A}_{\alpha} \big) &=&  -\frac{4}{q^{3} \bar{q}^{3}} \varepsilon^{\alpha\mu\lambda\nu}n_{\alpha} \big( q^{2} (\pa_{\lambda} A_{\mu})A_{\nu} + q \bar{q} (\pa_{\lambda} A_{\mu})V_{\nu} 
+\nonumber\\&& q \bar{q} (\pa_{\lambda} V_{\mu})A_{\nu} + \bar{q}^{2} (\pa_{\lambda} V_{\mu})V_{\nu}  \big).
\eea
Thus, we find the final result to Eq.(\ref{28}) in the following form
\bea \label{30}
 S^{(3)}_{eff} &=&\frac{1}{6\pi^2}  \int d^{4}x\ \Big(q^{2} n_{\alpha}\tilde{F}^{\alpha\nu} A_{\nu}+2q\tilde{q} \ n_{\alpha}n^{\sigma}\tilde{F}^{\alpha\nu}\tilde{F}_{\nu\sigma} +\nonumber\\
&& \tilde{q}^{2} n_{\alpha}n^{\sigma}n^{\theta}\varepsilon^{\mu\lambda\nu\alpha}(\pa_{\lambda}\tilde{F}_{\mu\sigma}) \tilde{F}_{\nu\theta}\Big),
\eea
which is finite in the limit $D\to 4$. 
{ Let us relate each term of Eq.(\ref{30}) with the effective action given by Eq.(\ref{livs}). Note that the first term of Eq.(\ref{30}) can be directly associated with the CPT-odd effective action, $S_{CS}$. The second term of Eq.(\ref{30}) can be recast as
\bea\label{31}
 \ n_{\alpha}n^{\sigma}\tilde{F}^{\alpha\nu}\tilde{F}_{\nu\sigma} \to -\frac{1}{2} n^{2} F_{\mu\nu} F^{\mu\nu} + n^{\mu}n_{\nu} F_{\alpha\mu} F^{\alpha\nu}.
\eea 
Note that the first term is similar to the usual Maxwell term. For the light-like four-vector $n_{\mu}=(n_{0}, {\bf n} )$, i.e., $n^{2}=0$, 
this contribution disappears.  The second term of Eq.(\ref{31}) can be related to CPT-even effective action, $S_{AET}$, that is, the aether-like Lorentz-violating term \cite{petrov2}. And finally, the third term of Eq.(\ref{30}) can also be rewriten as 
\bea\label{32}
n_{\alpha}n^{\sigma}n^{\theta}\varepsilon^{\mu\lambda\nu\alpha} (\pa_{\lambda}\tilde{F}_{\mu\sigma}) \tilde{F}_{\nu\theta}
\to  - \frac{1}{2} \big(  n^{\beta} F_{\lambda\beta}\big) (n\cdot\pa) \big( n_{\alpha} \tilde{F}^{\lambda\alpha}  \big) + \frac{1}{2}
n^{2}  \big(  \pa^{\beta} F_{\lambda\beta}\big)  \big(  n_{\alpha}\tilde{F}_{\lambda\alpha}\big).
\eea
The Eq.(\ref{32}) is similar to the full CPT-odd higher derivative operator  \cite{full} and in the light-like regime, $n^{2}=0$,
we recover the effective actions $S_{ECS}$, the electromagnetic Myers-Pospelov action. 
The radiatively induction of higher derivative terms has also been studied in \cite{petrov1}. 
}

\section{Conclusions}\label{sec04}
{ In summary, we deal with the fermion sector of a Lorentz-symmetry violating {\it extended} QED and rewrite it in terms of projection operators and effective fields. We consider this theory to radiatively induce general effective actions which show aspects of Lorentz violation in the electromagnetic sector. Differently of the Lorentz-violating radiative inductions studied in the literature, where each effective action is induced separately, in our main result, Eq.(\ref{30}), it is shown that the important Lorentz-violating terms such as  CPT-odd Chern-Simons-like action,  aether-like action and higher derivative action can be generated simultaneously. In this case, the factor $q^{2}/6\pi^{2}$ is a result that has also been obtained in the literature in an explicitly independent  regularization scheme \cite{afree}.  By identifying the right side of 
Eq.(\ref{31}) as a Lagrangian density of an non-dispersive dielectric medium, we can have the following correspondence $\varepsilon\sim n_0^2$ and $1/\mu\sim\vec{n}^2$, where $\varepsilon$ and $\mu$ are the dielectric and magnetic constants of the medium. Finally, our computations present new contributions to {\it extended} QED, which may find several applications that run from cosmological models to condensed matter systems. Such issues should be addressed elsewhere.
}

{\acknowledgments} We would like to thank to CNPq and CAPES for partial financial support.

\appendix

\section{}

\label{APPA}
The purpose of this appendix is to present some specific results associated with dimensional regularization scheme. In summary, the integrals are promoted to $D$ dimensions, that is, we change $\int d^{4}p/(2\pi)^{4}\to \big(\mu^{2}\big)^{\epsilon/2}\int d^{D}p/(2\pi)^{D}$ where $\epsilon=4-D$ and $\alpha(\epsilon)\equiv\left(\frac{4\pi\mu^{2}}{m^2}\right)^{\epsilon/2}$ is a real function of an arbitrary parameter $\mu^{2}$ that identifies the mass scale. Below we list several useful results:
\bea\label{In01}
\big(\mu^{2}\big)^{\epsilon/2}\int\frac{d^{D}p}{(2\pi)^{D}}\frac{1}{(p^{2} - m^{2})^{n}}&=&\frac{i \alpha(\epsilon)}{(4\pi)^2}\frac{(-1)^{n}}{\big(m^{2}\big)^{n - 2}}\frac{\Gamma(n -2 + \epsilon/2)}{\Gamma(n)}\nonumber\\ &\stackrel{n=2}{\rightarrow}& \frac{i\alpha(\epsilon)}{(4\pi)^2}\Gamma\Big(\frac{\epsilon}{2}\Big); 
\nonumber\\
&\stackrel{n=3}{\rightarrow}&-\frac{i \alpha(\epsilon)}{(4\pi)^2}\frac{1}{2 m^{2}}\Gamma\Big(1+\frac{\epsilon}{2}\Big)\nonumber\\&=&
-\frac{i \alpha(\epsilon)}{(4\pi)^2}\frac{1}{2 m^{2}}\frac{\epsilon}{2}\Gamma\Big(\frac{\epsilon}{2}\Big);
\eea
\bea\label{In02}
 \big(\mu^{2}\big)^{\epsilon/2}\int\frac{d^{D}p}{(2\pi)^{D}}\frac{p^{2}}{(p^{2} - m^{2})^{n}}
&=& \frac{i \alpha(\epsilon)}{(4\pi)^2}\frac{(-1)^{n-1}(4-\epsilon)}{\big(m^{2}\big)^{n - 3}}\frac{\Gamma(n -3 + \epsilon/2)}{2\Gamma(n)}\nonumber\\&
\stackrel{n=2}{\rightarrow}&  \frac{i\alpha(\epsilon)}{(4\pi)^2}\frac{(\epsilon - 4)}{2} m^{2}\Gamma\Big(-1+\frac{\epsilon}{2}\Big)\nonumber\\ &=&    -   \frac{i\alpha(\epsilon)}{(4\pi)^2}\frac{(\epsilon - 4)}{(\epsilon - 2)} m^{2}\,\Gamma\Big(\frac{\epsilon}{2}\Big);\nonumber\\
&\stackrel{n=3}{\rightarrow}& \frac{i \alpha(\epsilon)}{(4\pi)^2}\frac{(4-\epsilon)}{4} \Gamma\Big(\frac{\epsilon}{2}\Big);
\eea
\bea\label{In03}
 \big(\mu^{2}\big)^{\epsilon/2}\int\frac{d^{D}p}{(2\pi)^{D}}\frac{p^{\mu}p^{\nu}}{(p^{2} - m^{2})^{n}}
&=& \frac{i\alpha(\epsilon)}{(4\pi)^2}\frac{(-1)^{n-1}}{\big(m^{2}\big)^{n -3}}\frac{\Gamma(n -3 + \epsilon/2))}{2\Gamma(n)}\eta^{\mu\nu}\nonumber\\&\stackrel{n=2}{\rightarrow}&-\frac{i\alpha(\epsilon)}{(4\pi)^2}\frac{m^{2}}{ 2} \Gamma\Big(-1+\frac{\epsilon}{2}\Big)\eta^{\mu\nu}\nonumber\\ &=& \frac{i\alpha(\epsilon)}{(4\pi)^2}\frac{m^{2}}{  (\epsilon - 2)} \Gamma\Big(\frac{\epsilon}{2}\Big)\eta^{\mu\nu};\nonumber\\
&\stackrel{n=3}{\rightarrow}& \frac{i \alpha(\epsilon)}{(4\pi)^2}\frac{1}{4} \Gamma\Big(\frac{\epsilon}{2}\Big)\eta^{\mu\nu};\nonumber\\&\stackrel{n=4}{\rightarrow}&-\frac{i\alpha(\epsilon)}{(4\pi)^2}\frac{1}{12m^{2}} \Gamma\Big(1+\frac{\epsilon}{2}\Big)\eta^{\mu\nu}\nonumber \\&=& -\frac{i\alpha(\epsilon)}{(4\pi)^2}\frac{\epsilon}{24 m^{2}} \Gamma\Big(\frac{\epsilon}{2}\Big)\eta^{\mu\nu};
\eea
\bea\label{In04}
\big(\mu^{2}\big)^{\epsilon/2}\int\frac{d^{D}p}{(2\pi)^{D}}\frac{p_{\mu}p^{\mu}p^{\lambda}p_{\beta}}{(p^{2} - m^{2})^{n}}&=& \frac{i\alpha(\epsilon)}{(4\pi)^2}\frac{i(-1)^{n}(6-\epsilon)\,}{\big(m^{2}\big)^{n - 4}}\frac{\Gamma(n - 4+\epsilon/2)}{4\Gamma(n)}\delta^{\lambda}_{\beta}\nonumber\\&\stackrel{n=4}{\rightarrow}&\!\!\! \frac{i \alpha(\epsilon)}{(4\pi)^2}\frac{(6-\epsilon)}{24}\Gamma\Big(\frac{\epsilon}{2}\Big)\delta^{\lambda}_{\beta}.
\eea
and, finally 
\bea\label{In05}
\big(\mu^{2}\big)^{\epsilon/2}\int\frac{d^{D}p}{(2\pi)^{D}}\frac{p^{\mu}p^{\nu}p^{\lambda}p^{\alpha}}{(p^{2} - m^{2})^{n}}&=& \frac{i \alpha(\epsilon)}{(4\pi)^2}\frac{(-1)^{n}\,}{\big(m^{2}\big)^{n - 4}}\frac{\Gamma(n - 4+\epsilon/2)}{4\Gamma(n)}(\eta^{\alpha\nu}\eta^{\lambda\mu}+\eta^{\alpha\mu}\eta^{\lambda\nu}+\eta^{\alpha\lambda}\eta^{\mu\nu})\nonumber\\&\stackrel{n=4}{\rightarrow}&\!\!\! \frac{i \alpha(\epsilon)}{(4\pi)^2}\frac{1}{24}\Gamma\Big(\frac{\epsilon}{2}\Big)(\eta^{\alpha\nu}\eta^{\lambda\mu}+\eta^{\alpha\mu}\eta^{\lambda\nu}+\eta^{\alpha\lambda}\eta^{\mu\nu}).
\eea
Notice that to obtain the above results, we use the following expansions: 
\begin{subequations}
\bea\label{Rl01}
\alpha(\epsilon) = 1 + \frac{\epsilon}{2} {\rm log} \Big( \frac{4\pi \mu^{2}}{m^{2}}\Big) + {\cal O}(\epsilon^{2});
\eea
\bea\label{RI02}
\Gamma\Big(\frac{\epsilon}{2}\Big) =  \frac{2}{\epsilon} - \gamma + {\cal O}(\epsilon).
\eea
\end{subequations}
Here, $\gamma\approx 0.577$ is the Euler-Mascheroni constant.


\end{document}